\documentclass[onecolumn,showpacs,preprintnumbers,amsmath,amssymb]{revtex4}

\usepackage{graphicx}
\usepackage{amsmath}
\usepackage{bm}
\usepackage{multirow}
\usepackage{comment}
\newcommand{\simgt}{\lower.5ex\hbox{$\; \buildrel > \over \sim \;$}}
\newcommand{\simlt}{\lower.5ex\hbox{$\; \buildrel < \over \sim \;$}}

\newcommand{\et}{{\it et al}}

\def\bfk{{k,\mu}}

\def\tiP{{\widetilde{P}}}
\def\rmr{({\rm r})}
\def\rms{({\rm s})}
\def\rmlin{{\rm lin}}

\begin{document}
\title{%        %You can use \\ for explicit line-break
Confronting the damping of the baryon acoustic oscillations 
with observation}

%\subtitle{Subtitle}    %use this when you want a subtitle

\author{%       %Use \scshape  for the family name
Hidenori {Nomura}$^{1}$, Kazuhiro {Yamamoto}$^{1}$, 
Gert {H\"utsi}$^{2,3}$, and Takahiro Nishimichi$^4$}

\affiliation{%     %Affiliation, neglected when [addenda] or [errata]
$^{1}$Department of Physical Science, Hiroshima University,
Higashi-Hiroshima 739-8526,~Japan
\\
$^{2}$Department of Physics and Astronomy, University College London, London,
WC1E 6BT,~UK
\\
$^{3}$Tartu Observatory, EE-61602 T\~oravere, Estonia\\
$^{4}$Department of Physics, School of Science, The University of Tokyo, 
Tokyo 113-0033, Japan}

%\publishedin{%         %Write this ONLY in cases of addenda and errata
%Prog.~Theor.~Phys.\ \textbf{XX} (19YY), page.}

%\recdate{Mmmmm DD, YYYY}%            %Editorial Office will fill in this.

%\abst{%         %this abstract is neglected when [addenda] or [errata]
\begin{abstract}
We investigate the damping of the baryon acoustic oscillations 
in the matter power spectrum due to the quasinonlinear clustering and 
redshift-space distortions by confronting the models with the observations 
of the Sloan Digital Sky Survey luminous red galaxy sample. 
The chi-squared test suggests that the observed power spectrum is better
matched by models with the damping of the baryon acoustic oscillations 
rather than the ones without the damping.
\end{abstract}

\maketitle

%**********************************************************************
\section{Introduction}
%**********************************************************************
The baryon acoustic oscillations (BAO), the sound oscillations
of the primeval baryon-photon fluid prior to the recombination 
epoch, left their signature in the matter power spectrum 
\cite{EH,MWP}. 
The BAO signature in the galaxy clustering has recently attracted 
remarkable attention as a powerful probe for exploring the 
nature of the dark energy component commonly 
believed to be responsible for 
the accelerated expansion of the Universe \cite{Eisenstein,
Hutsi,PercivalI,PercivalII,Tegmark,Yamamoto03}.
The usefulness of the BAO to constrain the dark energy has been 
demonstrated \cite{PercivalIII,Okumura,HutsiII}, and 
a lot of the BAO survey projects are in progress
or planned \cite{sdss3,Bassett,lsst,ska,Robberto}.
The BAO signature in the matter clustering plays a role of the 
standard ruler, because the characteristic scale of the BAO is 
well understood within the cosmological linear perturbation theory
as long as the adiabatic initial density perturbation is assumed. 
%This is the reason why the BAO is useful as a probe of the dark energy, 
%which is quite analogous to the supernovae which play a role of 
%the standard candles for exploring the dark energy. 

However, the comparison of the BAO signature with observation is 
rather complicated. The observed galaxy power spectrum is 
contaminated by the nonlinear evolution of the density perturbations,
the redshift-space distortions and the clustering bias.
%\cite{The clustering bias is also, which is mentioned in summary.}.
This enables us to use the 
galaxy power spectrum for other supplementary tests, in addition to
the test of the expansion history of the Universe for the equation
of state of the dark energy. For example, the redshift-space
distortions probe the linear growth rate of the density 
fluctuations \cite{Guzzo,YSH,PS}. The growth rate is now 
recognized to be very important as the test of gravity on the
cosmological scales. 

In the paper \cite{Nomura}, some of the authors of the present paper
investigated how the quasinonlinear density perturbations affect
the BAO signature. Especially, we focused on the damping of 
the BAO signature.
%in the matter power spectrum due to the quasi-nonlinear clustering. 
The semianalytic investigation on the basis of 
the third-order perturbation theory demonstrated that the BAO damping
is sensitive to the growth factor $D_1(z)$ and the amplitude 
of the matter power spectrum $\sigma_8$. Here $z$ is the redshift and 
the growth factor is normalized 
as $D_1(z)=a$ at $a\ll1$, where $a$ is the scale factor normalized 
as $a=1$ at the present epoch.
As a result, a measurement of the BAO damping might be useful as an
additional consistency test by enabling one to probe
the growth factor multiplied by the amplitude of the matter 
perturbation $D_1(z)\sigma_8$. 
In the present paper, we extend the previous work to include the
redshift-space distortions, and confront the 
BAO damping with the observed SDSS LRG galaxy power spectrum.
Throughout this paper, we use units in which the velocity of light
equals 1, and adopt the Hubble parameter $H_0=100h {\rm km/s/Mpc}$ 
with $h=0.7$.

%**********************************************************************
\section{Damping of the BAO}
%{\it Damping of the BAO} --
%**************************************************************************
We start with reviewing the theoretical modeling of the BAO damping. 
The BAO signature is extracted from the matter power spectrum 
$P(\bfk,z)$ at redshift $z$ in the following manner,
\begin{eqnarray}
  B(\bfk,z)\equiv {P(\bfk,z)\over \tiP(\bfk,z)}-1,
\label{defb}
\end{eqnarray}
where $\mu$ is the cosine of the angle between the line of sight 
direction and the wave number vector, and $\tiP(\bfk,z)$ is the 
corresponding smooth spectrum without the BAO. 
As will be explained in detail below, we adopt the formalism 
developed by Matsubara \cite{Matsubara} %, which uses the 
%technique of resuming infinite series of higher order 
%perturbations on the basis of the Lagrangian perturbation theory, 
for theoretical modeling of $P(\bfk,z)$.   
The corresponding smooth spectrum $\tiP(\bfk,z)$ is computed
in the same manner as $P(\bfk,z)$ but with the no-wiggle 
transfer function in Ref.~\cite{EH}. %, which enables us to investigate the
%damping of the BAO in an analytic way \cite{Nomura}.
As an alternative method,
one can utilize the cubic spline fitting method to 
construct the smooth spectrum \cite{PercivalII,Nishimichi}, 
which we adopt in comparison with observations.
%{\bf This kind of the definition Eq.~(\ref{defb}) 
%enables us to investigate the
%damping of the BAO in an analytic way \cite{Nomura}.}

The modeling of the quasinonlinear power spectrum has been
investigated by many authors, based on both the perturbation theory 
and numerical simulations. As a nonperturbative approach beyond 
the standard perturbation theory, 
%(SPT),\cite{Juszkiewicz,Vishniac,Fry,Goroff,Makino,Bertschinger,Bernardeau}
Matsubara proposed a model of the quasinonlinear matter power
spectrum using the technique of resuming infinite series 
of higher order perturbations on the basis of the Lagrangian 
perturbation theory (LPT) \cite{Matsubara}.
One of the advantages of using the LPT framework is the ability to
calculate the quasinonlinear matter power spectrum in redshift space,
%
%In the LPT approach, the quasinonlinear matter power spectrum in redshift
%space, $P_{\rm LPT}^{\rms}(\bfk,z)$, 
which can be obtained by
\begin{eqnarray}
 &&\hspace{-3mm}P_{\rm LPT}^{\rms}(\bfk,z)=e^{-\alpha(\mu,z)D_1^2(z)g(k)}
  \left[D_1^2(z)P_{\rmlin}^{(s)}(\bfk)
%\right.
%\nonumber \\
%&&\left.
% \hspace{-3mm}
+D_1^4(z)P_{2}^{(s)}(\bfk)
  +\alpha(\mu,z)D_1^4(z)g(k)P_{\rmlin}^{(s)}(\bfk)\right],
\label{lpts}
\end{eqnarray}
where
\begin{eqnarray}
&&g(k)={k^2\over 6\pi^2}\int_0^\infty dq P_{\rmlin}^{\rmr}(q),
\\
&&P_{\rmlin}^{(s)}(\bfk)=\left(1+f\mu^2\right)^2P_{\rmlin}^{\rmr}(k),
\end{eqnarray}
%\begin{eqnarray}
% &&g(k)={k^2\over 6\pi^2}\int dq P_{\rmlin}^{\rmr}(q)
%\\
% &&P_{\rmlin}^{(s)}(\bfk)=\left(1+f\mu^2\right)^2P_{\rmlin}^{\rmr}(k),
%\label{power_depmu}
%\end{eqnarray}
$P_{\rmlin}^{\rmr}(k)$ is the linear matter power spectrum 
at the present epoch, $f=d\ln D_1/d\ln a$, and $\alpha(\mu,z)=1+f(f+2)\mu^2$.
Also, $P_2^{(s)}(k)$ is expressed as
\begin{eqnarray}
 P_2^{(s)}(\bfk)&=&P_{22}^{\rms}(\bfk)+P_{13}^{\rms}(\bfk),
\end{eqnarray}
where %the explicit expressions of $P_{22}^{\rms}(k)$ and
%x$P_{13}^{\rms}(k)$ are given by %Appendix A
\begin{eqnarray}
%\hspace{-11mm}
 &&\hspace{-3mm}%\left.
  P_{22}^{\rms}(\bfk)=\sum_{n,m}\mu^{2n}f^m{k^3\over 4\pi^2}
  \int_0^{\infty} dr P_{\rmlin}^{\rmr}(kr)
% \right.%\hspace{35mm}
%\nonumber \\
%  &&\hspace{-3mm}\times
\int_{-1}^1 dx P_{\rmlin}^{\rmr}[k(1+r^2-2rx)^{1/2}]
  {A_{nm}(r,x)\over (1+r^2-2rx)^2},
\\
 &&\hspace{-3mm}P_{13}^{\rms}(\bfk)=(1+f\mu^2)P_{\rmlin}^{\rmr}(k)
%\nonumber \\
%  &&\hspace{0.3cm}\times 
\sum_{n,m}
  \mu^{2n}f^m{k^3\over 4\pi^2}
  \int_0^{\infty} dr P_{\rmlin}^{\rmr}(kr)B_{nm}(r),
\end{eqnarray}
and $A_{nm}(r,x)$ and $B_{nm}(r)$ are given in Appendix B of
Ref.~\cite{Matsubara}. We take $P_{\rm LPT}^{\rms}(\bfk,z)$ as
$P(\bfk,z)$ in Eq.~(\ref{defb}).%, and  $\tiP(\bfk,z)$ is obtained
%in a similar way with the no-wiggle transfer function. 

%---------------------------------------------------------------
%\begin{figure}[htbp]
\begin{figure}[t]
 \includegraphics[width=120mm, height=120mm]{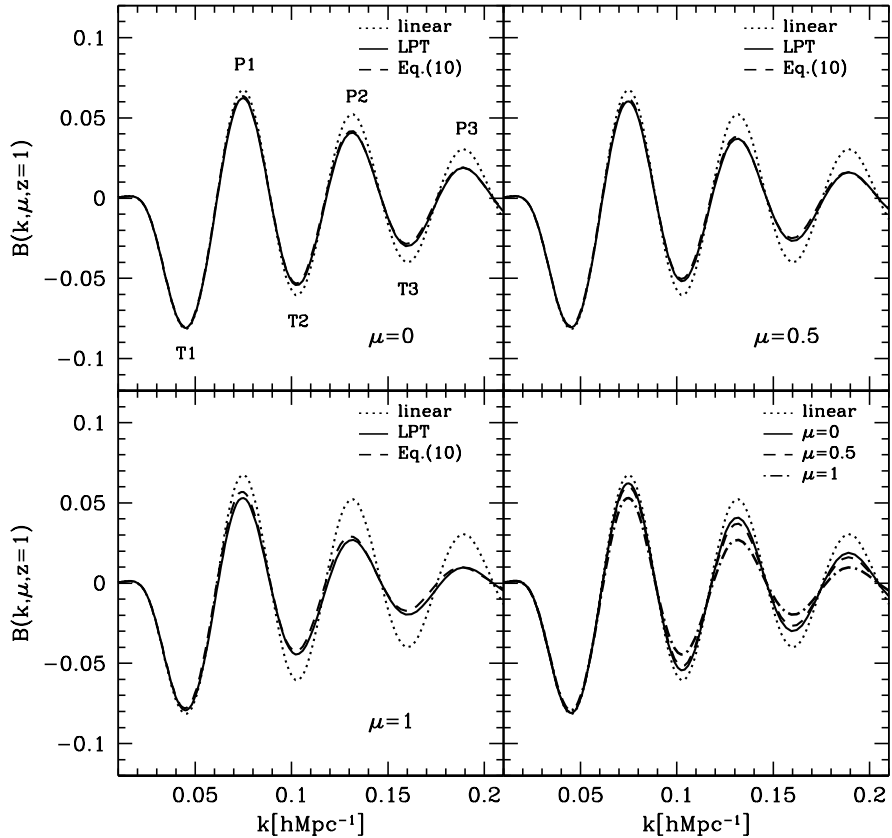}
 \caption{ The BAO signature according to the linear theory and the LPT 
 formalism.
 Except for the lower right panel, the dotted curve is the linear 
 theory, the solid curve is the LPT formula, and the dashed curve 
 uses the approximate formula Eq.~(\ref{w_lpt_red}), for 
 $\mu=0,~0.5$, and $1$, respectively. The lower right panel summarizes
 the $\mu$-dependence, obtained with Eq.~(\ref{bex}).
 The (quasinonlinear) redshift-space 
 distortion causes more damping of the BAO signature. 
 {Here the redshift is $z=1$, and the cosmological parameters are 
 $h=0.7$, $\Omega_m=0.27$, $\Omega_b=0.046$, $n_s=0.96$ and
 $\sigma_8=0.82$}}.
 \label{fig1}
\end{figure}

\begin{figure}[t]
 \includegraphics[width=120mm, height=120mm]{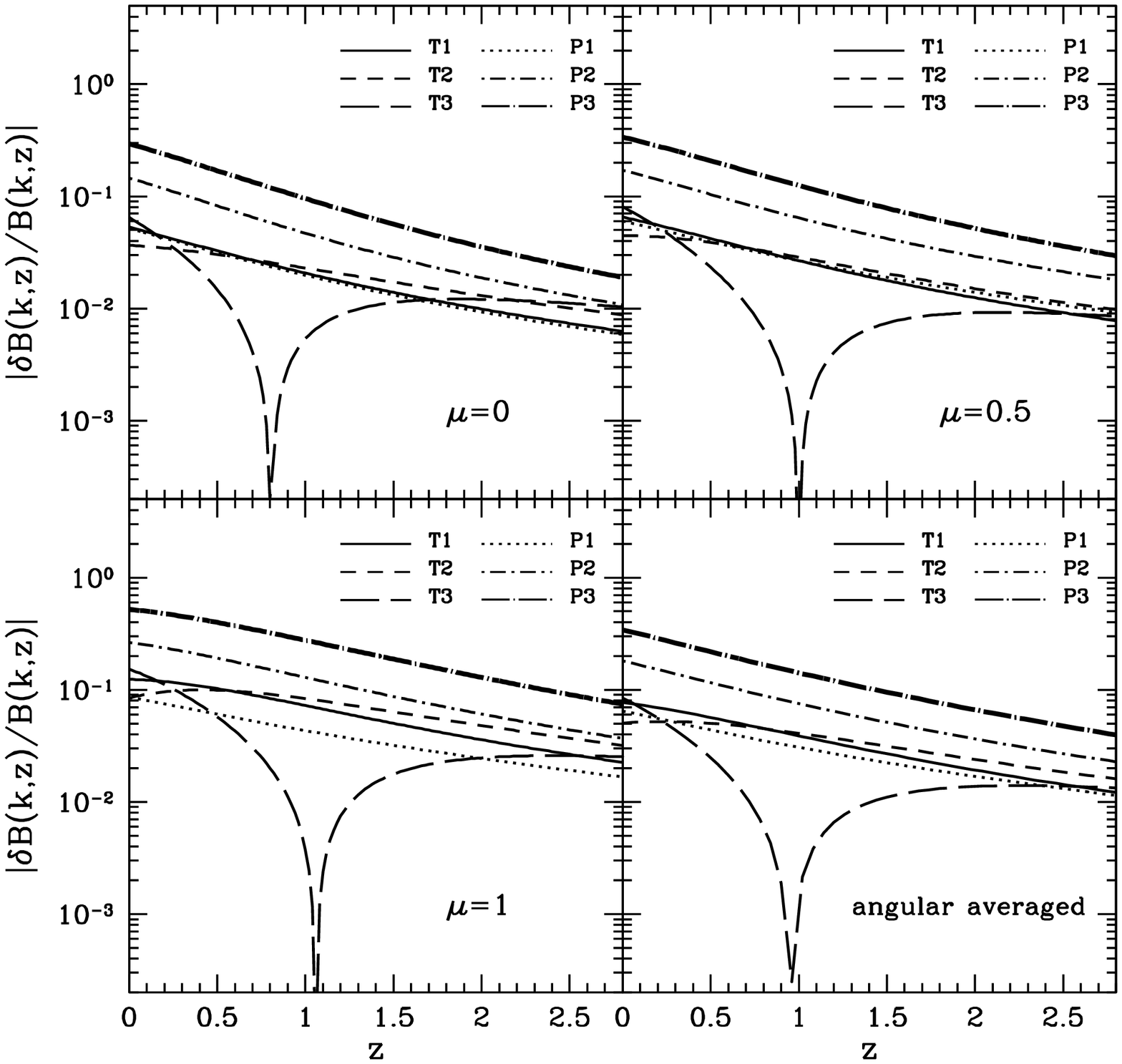}
 \caption{Relative error 
 $|\delta B/B|=|B_{\rm ex}-B_{\rm ap}|/|B_{\rm ex}|$
 at the wave numbers P1, P2, P3, T1, T2 and T3, 
 which correspond to the peaks and troughs defined in Fig.~1, as a function 
 of the redshift. Except for the right lower panel, the cases $\mu=0,~0.5$, 
and $1$
 are shown, respectively. The right lower panel shows the result for the angular
 averaged power spectrum  $|\delta B/B|=|B^{\rm ave}_{\rm ex}-B^{\rm ave}_{\rm ap}|
 /|B^{\rm ave}_{\rm ex}|$. }
 \label{fig2}
\end{figure}
%---------------------------------------------------------------

Figure 1 shows the BAO signature. Except for the right lower panel,
the dotted curve is the linear theory, while the solid curve is the result
from the LPT formula at redshift $z=1$ for $\mu=0,~0.5$, and $1$,
respectively, which is explicitly given by
\begin{eqnarray}
 B_{\rm ex}(\bfk,z)={P_{\rm LPT}^{\rms}(\bfk,z)
  \over \tiP_{\rm LPT}^{\rms}(\bfk,z)}-1.
\label{bex}
\end{eqnarray}
The right lower panel summarizes the $\mu$ dependence, 
which is given by Eq.~(\ref{bex}).
Note that the case $\mu=0$ is equivalent to the LPT formula in real 
space. 
%The cosmological parameters are 
%$\Omega_m=0.27$, $\Omega_b=0.046$, $n_s=0.96$ and  $\sigma_8=0.82$.
%
The amplitude of the BAO signature is degraded compared with 
the linear perturbation theory. % (dotted curve). 
Thus the quasinonlinear clustering and the
redshift-space distortions decrease the amplitude of the BAO. 

Let us introduce the correction function of the BAO damping 
$W(\bfk,z)$ by 
\begin{eqnarray}
 B_{\rm ap}(\bfk,z)=\left[1-W(\bfk,z)\right] B_{\rmlin}(k),
\label{bfitr}
\end{eqnarray}
where $B_{\rmlin}(k)$ is the BAO signature in linear theory.
In the previous paper \cite{Nomura}, which was restricted to 
real space, it was demonstrated that the correction factor 
can be written in a rather simple form.
One of the main results of the present paper is that a similar 
simple formula can be derived in redshift-space.
After some computation similar to the one in \cite{Nomura}, 
we found that the leading factor of the correction function 
can be approximately written as
\begin{eqnarray}
 W(\bfk,z)={D_1^2(z)\over 1+\alpha(\mu,z)D_1^2(z)\tilde{g}(k)}
  {\tiP_{22}^{\rms}(\bfk)\over \tiP_{\rmlin}^{\rms}(\bfk)},
\label{w_lpt_red}
\end{eqnarray}
where $\tiP_{22}^{\rms}(\bfk)$ and $\tiP_{\rmlin}^{\rms}(\bfk)$
are defined as $P_{22}^{\rms}(\bfk)$ and $P_{\rmlin}^{\rms}(\bfk)$, 
respectively, but
%{\bf without the BAO signature}.
with the no-wiggle transfer function.
The formula (\ref{w_lpt_red}) in the limit of $\mu=0$ reduces to the previous 
result derived for real space \cite{Nomura}. 
The dashed curve in Fig.~1 shows the approximate formula 
(\ref{bfitr}) with (\ref{w_lpt_red}). 
%In deriving Eq.~(\ref{w_lpt_red}), we adopted the approximate 
%formula
%\begin{eqnarray}
% P_{22}^{\rms}(k)&\simeq& \tiP_{22}^{\rms}(k).
%\\
% P_{13}^{\rms}(k)&\simeq& \left[1+B_{\rmlin}^{\rms}(k)\right]
%  \tiP_{13}^{\rmr}(k).
%\end{eqnarray}

To demonstrate the validity of the approximate formula (\ref{w_lpt_red}), 
Fig.~2 shows the relative error $|B_{\rm ex}-B_{\rm ap}|/|B_{\rm ex}|$
at wave numbers P1, P2, P3, T1, T2 and T3, which correspond to 
the peaks and troughs defined in Fig.~1, as a function of the redshift. 
The upper left panel is $\mu=0$, the upper right panel is $\mu=0.5$,
and the lower left panel is $\mu=1$, respectively. The lower right 
panel is the result for the angular averaged power spectrum 
$|B^{\rm ave}_{\rm ex}-B^{\rm ave}_{\rm ap}|/|B^{\rm ave}_{\rm ex}|$ 
(see below for details).
The approximate formula works at the 10 \% level.   

Figure~3 shows the correction function $W(\bfk,z)$ as a function
of the wave number $k$ at redshift $z=1$ for $\mu=0,~0.5$, and $1$, 
respectively. It is obtained with the approximate formula 
(\ref{w_lpt_red}). 
The dotted-dashed curve is the correction function for the 
angular averaged power spectrum $W^{\rm ave}(k,z)$ (see below). 
Thus the BAO damping due to the redshift-space distortion 
is more efficient compared to the result in real space. 

In practice, the angular averaged power spectrum is used 
in measuring the BAO signature, which is expressed, as follows,
using the power spectrum in the LPT formula:
\begin{eqnarray}
 B^{\rm ave}_{\rm ex}(k,z)&=&{\int_{-1}^1 d\mu P_{\rm LPT}^{(s)}(k,\mu,z)\over
  \int_{-1}^1 d\mu \tiP_{\rm LPT}^{(s)}(k,\mu,z)}-1.
\end{eqnarray}
With the use of Eqs.~(\ref{defb}) and (\ref{bfitr}), we find that
$B^{\rm ave}_{\rm ex}(k,z)$ is approximately written as 
\begin{eqnarray}
 B^{\rm ave}_{\rm ap}(k,z)&=&\left[1-W^{\rm ave}(k,z)\right]B_{\rm lin}(k),
\label{ave2}
\end{eqnarray}
with
\begin{eqnarray}
 W^{\rm ave}(k,z)={\int_{-1}^1 d\mu W(k,\mu,z)
 \tiP(k,\mu,z)\over \int_{-1}^1 d\mu \tiP(k,\mu,z)}.
\end{eqnarray}
The lower right panel of Fig.~2 shows the relative error 
$|B^{\rm ave}_{\rm ex}-B^{\rm ave}_{\rm ap}|/|B^{\rm ave}_{\rm ex}|$
as a function of the redshift at the wave numbers P1, P2, P3, T1, T2 and T3, 
which correspond to the peaks and troughs of the BAO.
The dotted-dashed curve in Fig.~3 plots $W^{\rm ave}(k,z)$ as a
function of $k$ at the redshift 1. 

Figure~4 compares the theoretical prediction
of the LPT formula with the results from the $N$-body 
simulations ($30$ realizations).
Each of our simulations used $512^3$ particles in periodic cubes 
with side length $10^3$ $h^{-1}$Mpc \cite{Nishimichinew}.
We apply a method to correct the deviation from the ideal case of 
infinite volume (see \cite{Nishimichinew} for details).  
The panels correspond to redshifts $z=3,2,1$, and $0.5$, 
respectively. 
%Due to the statistical limitation of the 
%$N$-body simulation, the agreement is not quite clear. 
One can see the agreement between the $N$-body result
and the theoretical prediction.
%for the larger wave numbers at lower redshifts.

%-------------------------------------------------------
\begin{figure}[htbp]
\includegraphics[width=90mm, height=90mm]{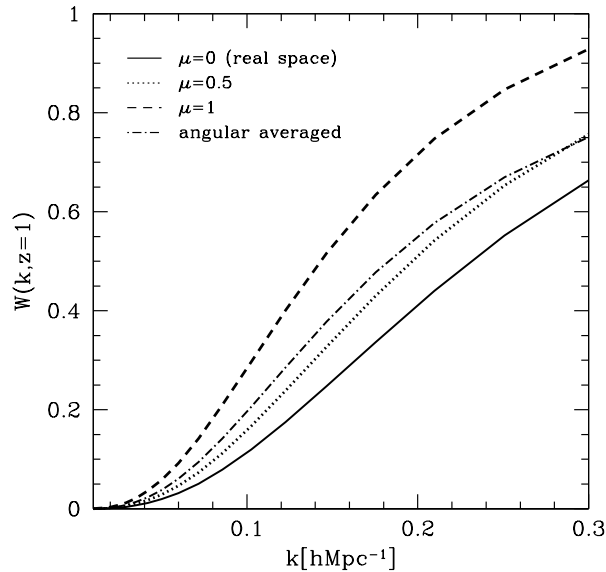}
\caption{The damping correction $W(\bfk,z)$ at redshift $z=1$ 
as a function of the wave number, for $\mu=0,~0.5$, and $1$, respectively. 
The dotted-dashed curve is the angular averaged case $W^{\rm ave}(k,z)$.
Here the cosmological parameters are the same as those of Fig.~1.}
\label{fig3}
\end{figure}
%\begin{figure}[htbp]
\begin{figure}[t]
\includegraphics[width=120mm, height=120mm]{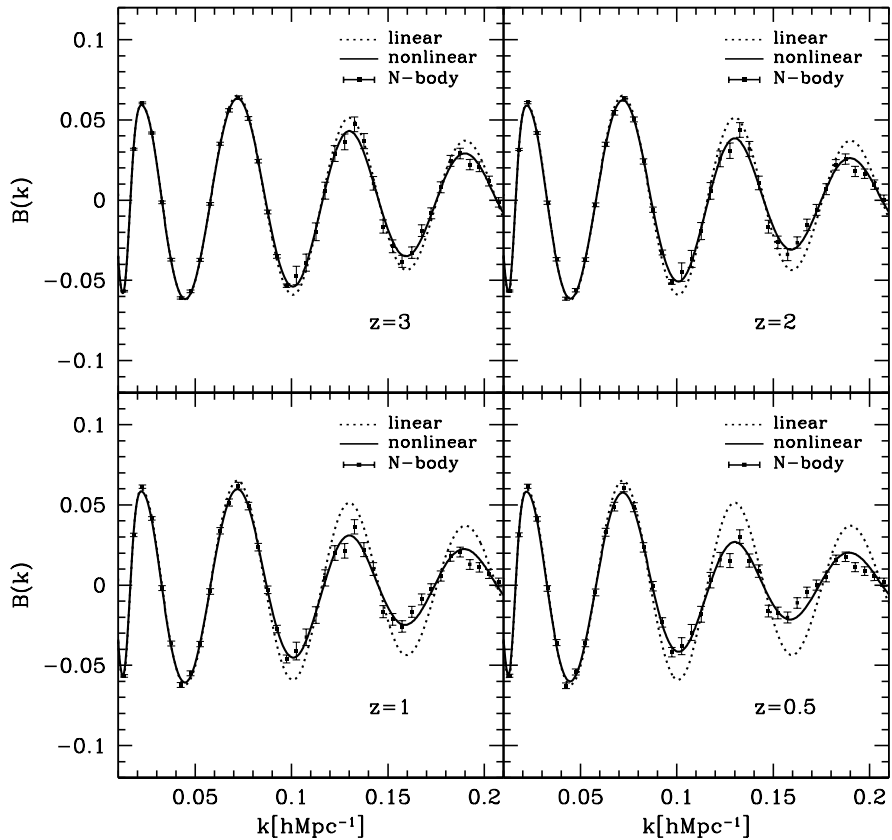}
\caption{Comparisons between the theoretical BAO signature 
and the results from the $N$-body simulation 
(squares with error bars) at $z=3,~2,~1$, and $0.5$, 
respectively. The solid curve is the LPT formula and the
dotted curve is the linear theory. The  solid line is the BAO distorted by
the nonlinear redshift-space distortions. The cosmological parameters
adopted in this comparison are
$h=0.7$, $\Omega_m=0.28$, $\Omega_b=0.046$, $n_s=0.96$
and  $\sigma_8=0.82$}
\label{fig4}
\end{figure}
%-------------------------------------------------------

%-------------------------------------------------------
\begin{figure}[htbp]
\includegraphics[width=90mm, height=90mm]{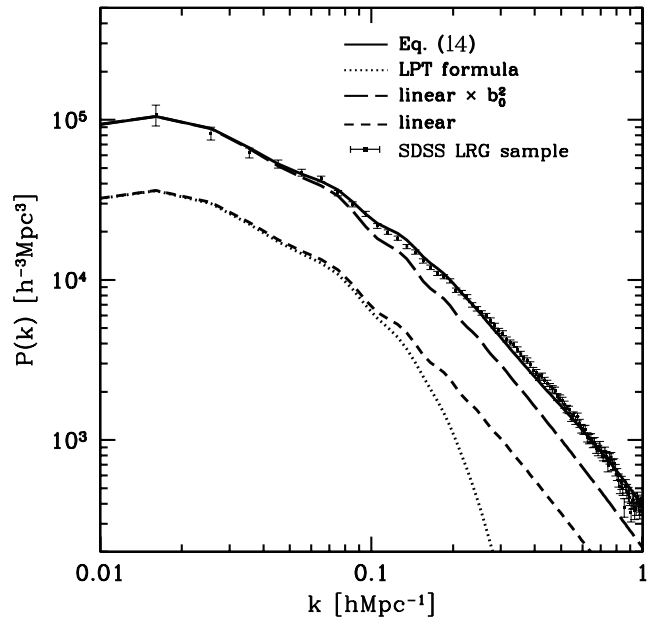}
\caption{Comparison of the theoretical power spectra and  
the SDSS LRG power spectrum.% (the square with the error bar). 
 The dotted curve is
the LPT formula, and the short-dashed curve is the linear theory.
The solid curve is the power spectrum multiplied by the
correction factor (\ref{fitsdss}).
%where $b_0=?$ and $\alpha=?$. 
The long-dashed curve is the linear theory multiplied by 
the constant $b_0^2$. 
The cosmological and fitting parameters are described in
 Table~I labelled as model no.1.}
\label{fig5}
\end{figure}
%-------------------------------------------------------
%-------------------------------------------------------
\begin{figure}[htbp]
\includegraphics[width=90mm, height=90mm]{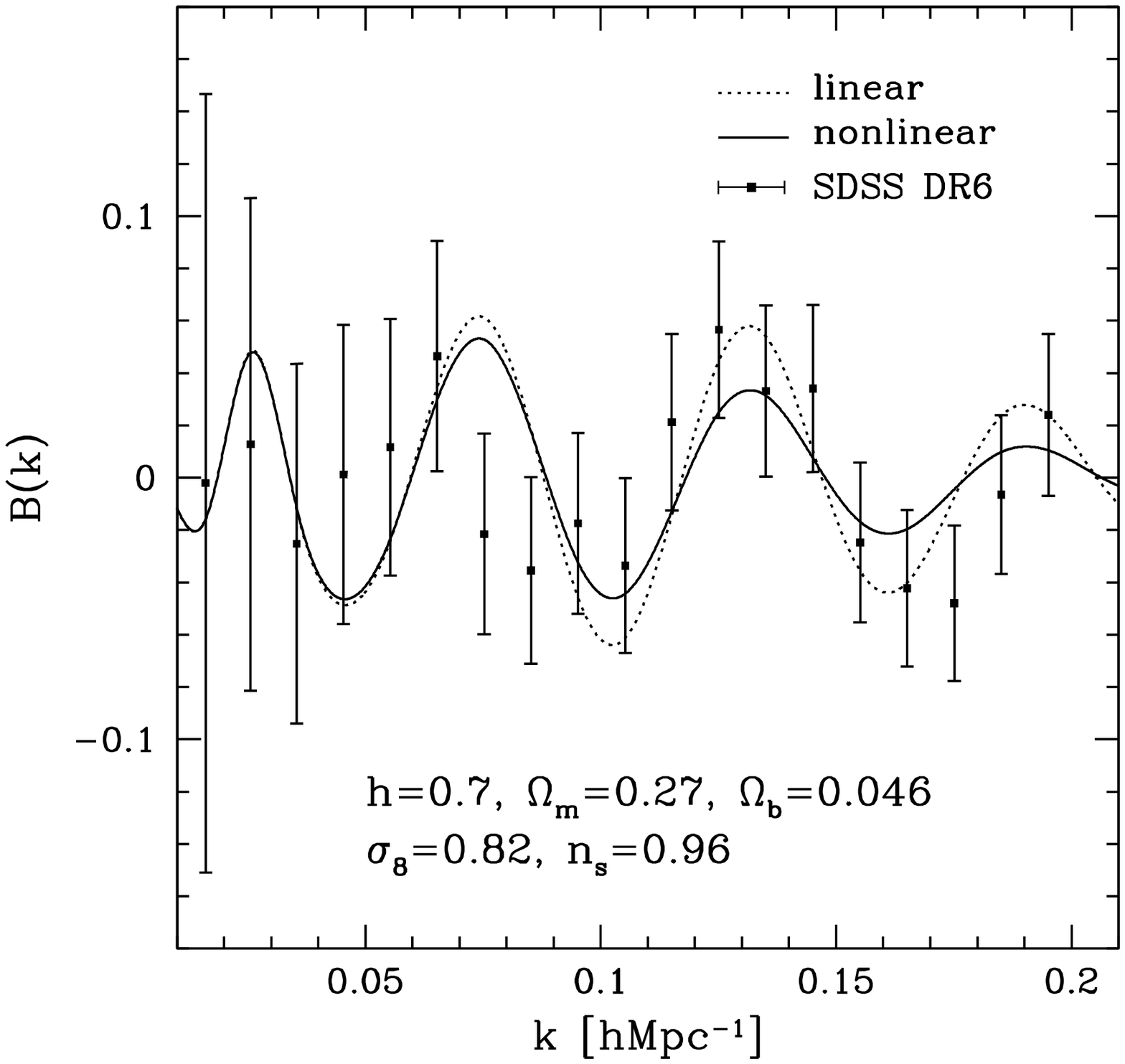}
\caption{Comparison of the BAO features extracted from 
theoretical models and from the observational data. 
The dotted curve is the linear theory and 
the solid curve is the LPT result. 
The squares with the error bars are the results from the SDSS LRG sample. 
The chi-squared score for this example is listed in 
Table~I, labelled as model no.1.
}
\label{fig6}
\end{figure}
%-------------------------------------------------------

%**********************************************************************
\section{Comparison with the SDSS LRG power spectrum}
%{\it Confrontation with the SDSS LRG spectrum} --
%**************************************************************************
Now we confront the theoretical predictions with observations.
In particular, we use the SDSS LRG sample from data release 6. 
The SDSS data reduction procedure is the same as described in 
Ref.~\cite{Hutsi}.
Here we utilize the cubic spline fit to {\it consistently} construct 
the smooth component for 
both the theoretical and the observational power spectra.
However, the overall shape of the power spectrum from the LPT formula
does not match the observational power spectrum. 
The power spectrum of the LPT formula shows the exponential 
suppression at large wave numbers, as is shown in Eq.~(\ref{lpts}).
This feature can be understood as the nonlinear redshift-space 
distortion  \cite{Matsubara},  the so-called finger-of-God effect,
which is automatically formulated in the LPT formalism.
This suppression factor matches a phenomenological model 
of the redshift-space power spectrum on large scales 
in Ref.~\cite{ESW}. 
Such discrepancies probably arise from the truncation of higher
order perturbations and ignoring the effect of galaxy clustering
bias. This might make a systematic error in extracting the 
BAO consistently.
To avoid this, we first construct the theoretical power spectrum 
by multiplying the LPT power spectrum by the function 
$b_0^2e^{\alpha k^2}$ so as to match the SDSS LRG power spectrum, 
\begin{eqnarray}
  P_{\rm fit}(k)=b_0^2e^{ \alpha k^2}P_{\rm LPT}^{\rms}(k),
\label{fitsdss}
\end{eqnarray}
where $b_0$ and $\alpha$ are the fitting parameters. 
Figure~5 demonstrates the example $P_{\rm fit}(k)$,
whose parameters are described in Table~I, labeled 
as model no.1.  One can see that this fitting 
function matches the observed power spectrum well.
We also note that the BAO signature extracted using 
$P_{\rm fit}(k)$ is not sensitive to the choice of 
$b_0$ and $\alpha$. 

\begin{table}[t]
\begin{center}
\begin{eqnarray}
{\footnotesize
\begin{array}{c@{\hspace{3.3mm}}c@{\hspace{3.3mm}}c
 @{\hspace{3.3mm}}c@{\hspace{3.3mm}}c@{\hspace{3.3mm}}c@{\hspace{3.3mm}}
  c@{\hspace{3.3mm}}c@{\hspace{3.3mm}}c@{\hspace{3.3mm}}
  c@{\hspace{3.3mm}}c@{\hspace{3.3mm}}c}\hline\hline
 {\rm No.}& \Omega_m & \Omega_b & \sigma_8 & n_s & b_0
 & \alpha[h^{-2}{\rm Mpc}^2]
 & \chi^2_{\rm lin}
 & \chi^2_{\rm LPT}
 & \chi^2_{\rm simple}
 & \chi^{2}_{\rm lin-cov}
 & \chi^{2}_{\rm LPT-cov}
\\ \hline
 1 & 0.27 & 0.046 & 0.82 & 0.96 & 1.7  & 27.5 & 15.1 & 13.7 & 13.7
 & 28.2 & 23.0
 \\[-2mm]
 2 & 0.27 & 0.048 & 0.82 & 0.96 & 1.7  & 27.2 & 13.9 & 13.0 & 12.9
 & 26.4 & 22.1
 \\[-2mm]
 3 & 0.27 & 0.044 & 0.82 & 0.96 & 1.7  & 27.8 & 16.6 & 14.5 & 14.6
 & 30.5 & 24.1
  \\[-2mm]
 4 & 0.27 & 0.046 & 0.80 & 0.96 & 1.75 & 26.2 & 15.1 & 13.7 & 13.7
 & 28.2 & 23.1
  \\[-2mm]
 5 & 0.27 & 0.046 & 0.84 & 0.96 & 1.65 & 28.8 & 15.1 & 13.7 & 13.7
 & 28.2 & 22.9
  \\[-2mm]
 6 & 0.27 & 0.046 & 0.82 & 0.94 & 1.7  & 28.0 & 14.9 & 13.5 & 13.5
 & 28.0 & 22.7
  \\[-2mm]
 7 & 0.27 & 0.046 & 0.82 & 0.98 & 1.7  & 27.0 & 15.2 & 13.8 & 13.8
 & 28.4 & 23.2
 \\[-2mm]
 8 & 0.26 & 0.046 & 0.82 & 0.96 & 1.65 & 28.8 & 15.3 & 12.5 & 13.0
 & 29.7 & 22.3
  \\[-2mm]
 9 & 0.28 & 0.046 & 0.82 & 0.96 & 1.75 & 26.0 & 15.3 & 15.0 & 14.5
 & 27.3 & 23.8
  \\
\hline
\end{array}
}
\nonumber
\end{eqnarray}
\caption{The results of the chi-squared test for the BAO signature for various 
cosmological models. $\chi^2_{\rm LPT}$ is based on the LPT power 
spectrum, while $\chi^2_{\rm lin}$ assumes the linear power spectrum.
$\chi^2_{\rm simple}$ is the minimum chi-squared value
in fitting the model $B(k)=[1-d_*^2k^2]B_{\rm lin}(k)$.
$\chi^{2}_{\rm lin-cov}$ and $\chi^{2}_{\rm LPT-cov}$ assume the linear
 and LPT power spectra, respectively, in evaluating Eq.~(\ref{chi_cov}).}
\end{center}
\end{table}

Figure~6 compares the BAO signatures extracted from the 
theoretical models and the SDSS LRG power spectrum of Fig.~5. 
We computed the chi-square as 
\begin{eqnarray}
\chi^2=\sum_{i}{[B^{\rm th}(k_i)-B^{\rm ob}(k_i)]^2
\over 
\Delta B(k_i)^2},
\end{eqnarray}
where $B^{\rm th}(k_i)$ and $B^{\rm ob}(k_i)$ are the theoretical and 
observational BAO signatures at wave number $k_i$, 
respectively, and $\Delta B(k_i)$ is the error. In the computation, 
we used the data in the wave number range of 
$0.015\leq k\leq 0.195$. 
The values of the chi-squared test for various cosmological models are 
listed in Table~I.
Here $\chi^2_{\rm LPT}$ is the result
for the theoretical LPT model, while $\chi^2_{\rm lin}$
is that for the linear theory, which does not take the BAO damping
into account.
In this computation, we have not fitted any parameters, and 
the number of degrees of freedom is $19$.
$\chi^2_{\rm LPT}<\chi^2_{\rm lin}$ for all of the models. 
This means that the models with the BAO damping match
the observational results better. 

As an additional test, we compared the observational BAO signature 
with a very simple theoretical model  
$B(k)=[1-d_*^2k^2]B_{\rm lin}(k)$, 
which includes the leading correction to the damping.
Taking $d_*$ as a free parameter, we computed the chi-square
and we found a minimum value at $d_{*}\simeq 4 h^{-1}{\rm Mpc}$, 
for the models in Table~I. 
$\chi^2_{\rm simple}$ is the minimum chi-square. 
Note that the case $d_*=0$ corresponds to the linear theory. 
Then, $\Delta \chi^2=\chi^2_{\rm lin}-\chi^2_{\rm simple}\sim1$,
which suggests that the detection of the BAO damping is at the 1 sigma level.

To see the effect of the covariance between the data points, we
compute
\begin{eqnarray}
&& \chi_{\rm cov}^2=\sum_{i,j}\left[B^{\rm th}(k_i)-B^{\rm ob}(k_i)\right]
  \tilde{P}(k_i){\rm Cov}^{-1}(k_i,k_j)\tilde{P}(k_j)
   \left[B^{\rm th}(k_j)-B^{\rm ob}(k_j)\right],
\label{chi_cov}
\end{eqnarray}
where ${\rm Cov}(k_i,k_j)$ is the covariance matrix of the power spectrum.
Here the covariance matrix is obtained by using $100$ mock catalogs generated
via the second-order Lagrangian perturbation theory and
Poisson sampling. The details of the procedure are described in 
Ref.~\cite{Hutsi}.  Figure~7 shows the resulting correlation 
matrix, which is defined by 
\begin{eqnarray}
{\rm r}(k_i,k_j)={{\rm Cov}(k_i,k_j)\over \sqrt {{\rm Cov}(k_i,k_i){\rm Cov}(k_j,k_j)}}.
\label{corrmat}
\end{eqnarray}
The result of $\chi^2_{\rm cov}$ value is shown in Table~I, 
where $\chi_{\rm LPT-cov}^{\rm }{}^2$ ($\chi_{\rm lin-cov}^{\rm }{}^2$) is the
result for the theoretical LPT model (the linear theory).
We find $\chi^2_{\rm LPT-cov}<\chi^2_{\rm lin-cov}$ for all the models, 
again. 

%-------------------------------------------------------
\begin{figure}[htbp]
\includegraphics[width=110mm, height=90mm]{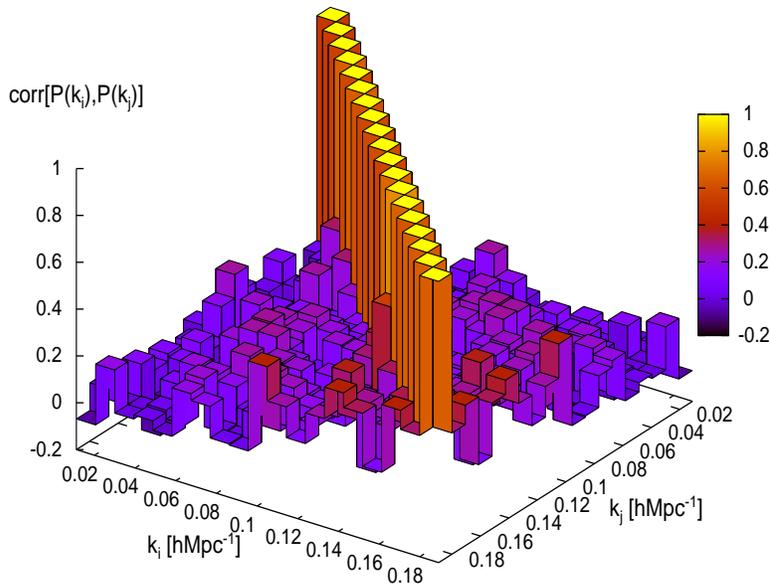}
\caption{Correlation matrix \cite{Hutsi}
obtained from 100 mock catalogs. 
}
\label{fig6}
\end{figure}
%-------------------------------------------------------

As discussed in Ref.\cite{Nomura}, to the leading order, the magnitude of the 
BAO damping is proportional to the amplitude of the matter power spectrum, 
$D_1(z)\sigma_8$. 
Thus, the precise measurement of the BAO damping might be useful in 
determining $D_1(z)\sigma_8$.
To estimate the minimum achievable error we have computed the diagonal 
entry for $D_1(z)\sigma_8$ in the inverse Fisher matrix. For the Fisher 
matrix calculation we have adopted the same approach as described in 
Ref.~\cite{Nomura} but here used the angular average of
$P(\bfk,z)$ instead of real space power spectrum. 
The results are almost the same as those for real space in 
Ref.~\cite{Nomura}. The minimum attainable error of $D_1(z=0.9)\sigma_8$ 
is $\sim 0.1~\times\left(\Delta A/2000 {\rm deg.}^2\right)^{1/2}$
(at the $1$ sigma level), where $\Delta A$ denotes the survey area. In
this computation, we assume that the galaxy sample covers the redshift
range $0.5\leq z\leq 1.3$, the mean number density of galaxies
$n=5.0\times 10^{-4}h^3{\rm Mpc}^{-3}$, and the clustering bias $b=2.0$.
Note that the error on $D_1(z=0.9)\sigma_8$ depends on the mean number
density of galaxies and the clustering bias \cite{Nomura}.

%**********************************************************************
\section{Summary and Conclusions}
%{\it Summary and Conclusions} -- 
%**************************************************************************
In the present work we investigated the influence of the 
redshift-space distortions 
on the damping of the BAO in the matter power spectrum.
The modeling was based on the work developed by Matsubara,
which uses the technique of resuming infinite series of higher 
order perturbations within the framework of the Lagrangian 
perturbation 
theory \cite{Matsubara}. The result shows that additional 
BAO damping appears due to redshift-space distortions. 
We confronted the theoretical BAO signature with the 
observed power spectrum of the SDSS LRG sample.
The chi-squared test suggests that the observed power spectrum
favors models with the BAO damping over the ones 
without the damping. Though the statistical significance 
is not high, the BAO damping has likely been detected in 
the SDSS LRG power spectrum.
In our modeling we have not taken into account the effect of 
the clustering
bias on the BAO damping. This should be considered more 
carefully (cf. \cite{MatsubaraII}); however, the authors of 
Ref.~\cite{SBA} show that the BAO damping does not depend 
much on the halo bias in redshift space.

%*************************************************************************
{\it Acknowledgements}
%*************************************************************************
This work was supported by a Grant-in-Aid 
for Scientific research of Japanese Ministry of Education, 
Culture, Sports, Science and Technology (No.~18540277)
TN is supported by a Grant-in-Aid from JSPS (No.DC1:~19-7066).

\end{document}